\documentclass[sigconf]{acmart}
\AtBeginDocument{%
  }

\usepackage{subcaption}
\usepackage{multirow} 

\newcommand{\UPZones}{urban preference zones}
\newcommand{\UPZonesTitle}{Urban Preference Zones}
\newcommand{\UPZonesShort}{UPZone}

\setcopyright{acmlicensed}
\copyrightyear{2025}
\acmYear{2025}
\acmDOI{XXXXXXX.XXXXXXX}
\acmConference[KDD '25]{30th ACM SIGKDD Conference on Knowledge Discovery and Data Mining}{August 03--07, 2025}{Toronto, Canada}
\acmISBN{978-1-4503-XXXX-X/2018/06}

\begin{document}

\title[Interest Networks (iNETs) for Cities]{Interest Networks (iNETs) for Cities: \\Cross-Platform Insights and Urban Behavior Explanations}


\author{Gustavo H. Santos}
\affiliation{%
  \small
 \institution{Universidade Tecnológica Federal do Paraná}
  \city{Curitiba}
  \country{Brazil}
}
\email{gustavohenriquesantos@alunos.utfpr.edu.br}

\author{Myriam Delgado}
\affiliation{%
  \small
  \institution{Universidade Tecnológica Federal do Paraná}
  \city{Curitiba}
  \country{Brazil}}
\email{myriamdelg@utfpr.edu.br}

\author{Thiago H. Silva}
\affiliation{%
  \small
  \institution{Universidade Tecnológica Federal do Paraná}
  \city{Curitiba}
  \country{Brazil}}
\email{thiagoh@utfpr.edu.br}

\renewcommand{\shortauthors}{Gustavo et al.}

\begin{abstract}

Location-Based Social Networks (LBSNs) provide a rich foundation for modeling urban behavior through iNETs (Interest Networks), which capture how user interests are distributed throughout urban spaces. This study compares iNETs across platforms (Google Places and Foursquare) and spatial granularities, showing that coarser levels reveal more consistent cross-platform patterns, while finer granularities expose subtle, platform-specific behaviors. Our analysis finds that, in general, user interest is primarily shaped by geographic proximity and venue similarity, while socioeconomic and political contexts play a lesser role. Building on these insights, we develop a multi-level, explainable recommendation system that predicts high-interest urban regions for different user types. The model adapts to behavior profiles---such as \textit{explorers}, who are driven by proximity, and \textit{returners}, who prefer familiar venues---and provides natural-language explanations using explainable AI (XAI) techniques. To support our approach, we introduce \texttt{h3-cities}, a tool for multi-scale spatial analysis, and release a public demo for interactively exploring personalized urban recommendations. Our findings contribute to urban mobility research by providing scalable, context-aware, and interpretable recommendation systems.

\end{abstract}



\received{15 May 2025}
\received[revised]{4 Jun 2025}
\received[accepted]{4 Jun 2025}

\maketitle

\section{Introduction}

\label{sec:intro}

Informative urban data obtained by Location-Based Social Networks (LBSNs) can reflect user preferences and mobility behaviors~\citep{10.1145/3323503.3360620,yang2019exploring,SILVER2023104130,liu2018comparing,silva2024using,Silva2019,chen2024_HumanMobility}. By capturing geolocated activities such as check-ins and reviews, these platforms enable the construction of Interest Networks (iNETs)—undirected graphs where nodes represent urban areas (e.g cities, neighborhoods, ZIP codes, etc...) and edges link areas co-visited by the same users. Our proposed iNETs allow exploring urban dynamics and can be used in applications such as recommendation systems, mobility prediction, and policy design. Our research comprises three phases.

\textbf{Formation of iNETs}:
We construct Interest Networks (iNETs) from user activity data drawn from LBSNs. These iNETs are built by connecting spatial regions whenever the same users interact with places in multiple areas. To support more comprehensive and interpretable analyses, we enrich each region with a set of contextual attributes, including socioeconomic indicators, cultural profiles, political leanings, and geographic features. This enriched representation allows for deeper investigation into the factors shaping user mobility, regional interest patterns, and cross-platform consistency.

\textbf{Validation across granularity levels and platforms}:
We systematically evaluate how iNETs perform across varying spatial granularities and urban configurations. Our results reveal that coarser levels of granularity lead to greater alignment between iNETs across platforms, while finer levels highlight subtle behavioral differences. To reconcile these discrepancies, we propose clustering high user interest - such clusters (namely, Urban Preference Zones - UPZones) appear more consistently across different platforms and spatial scales. The clusters offer a stable basis for comparison and act as interpretable reference points for further modeling.

\textbf{Understanding user interests}:
We leverage the validated iNETs to model user interests at both aggregated and individual levels. At the aggregated level, we analyze iNET structures to assess general mobility patterns. Our findings indicate that geographic proximity and venue similarity modestly influence user movement between areas, while socioeconomic and political factors show limited impact. At the individual level, we distinguish between high- and low-interest regions based on user behavior. In addition to general mobility traits, we examine returner vs. explorer behavioral patterns~\citep{Pappalardo2015}, and introduce a personalized multi-level recommendation model grounded in users' previously visited high-interest areas. Explainability is a key feature of this model: we apply XAI techniques to produce interpretable, natural-language explanations that connect individual preferences to spatial and behavioral characteristics. A demo showcasing this capability is available online\footnote{\url{https://cityhood.vercel.app/}}.


\section{Related Works}
\label{secRelated}

Prior studies have examined the complementarity of LBSN data, comparing platforms like Instagram and Foursquare \citep{silvaUrbcomp13}, proposing frameworks for urban dynamics \citep{marti2019social}, and exploring their role in neighborhood and tourism studies \citep{nolasco2022social}. While some validate LBSN data against official sources (e.g., World Values Survey\footnote{https://www.worldvaluessurvey.org}) \citep{SILVA201795}, our work focuses on how different platforms independently model urban behavior and where discrepancies emerge. Building on multi-scale approaches such as micro–meso–macro frameworks \citep{rogov_urban_2018} and wavelet transformations \citep{wu_multi-scale_2020}, we examine how granularity influences urban preference modeling. Additionally, while prior work has used clustering techniques on LBSN and POI data to define urban zones, such as Livehoods \citep{cranshaw_livehoods_2012} and functional topic models \citep{gao_extracting_2017}, our grid-based method aggregates spatial cells based on user behavior to ensure both structural and behavioral consistency.

Research on urban mobility has leveraged large-scale data to uncover patterns such as Lévy flight dynamics from Twitter traces \cite{Cheng2021}, migration flows through temporal graphs \cite{Huang2023}, class-based transit differences using smart card data~\cite{su12229603}, and formulate models to explain urban change \citep{SILVER2023104130}. Moving beyond trajectory modeling, our work maps user interests across cities and neighborhoods. 
We extend the returner/explorer framework~\cite{Pappalardo2015} by integrating contextual, geographic, and cultural factors into an explainable, multi-scale recommendation system that predicts urban interest and justifies suggestions via user behavior and regional signals—advancing transparency beyond POI-level systems~\cite{cheng2012fused,9036967,de2024explainable}.

\section{Data}

\subsection{LBSN Datasets}
\label{sec:LBSNs}

\noindent \textbf{Foursquare} (FS): We use public check-in data shared on Twitter, covering different cities in different continents. Each dataset entry, curated by \cite{SILVA201795}, includes check-in time, venue name/category, and user ID, for a total of 7,537,057 check-ins made by 1,403,808 users on 2,739,548 venues. 

\noindent \textbf{Google Places} (GP): This dataset, provided by \cite{He2017,Pasricha2018} for academic use, contains 11,453,845 reviews from 5,054,567 users of Google Plus for 3,114,353 venues on Google Maps around the world. It includes user details (name, education, employment), review text (multilingual), ratings, timestamps, and unique user IDs, along with venue information (name, category, hours, contact, address, coordinates).

\noindent \textbf{Extended Google Places} (GP+): We also consider a bigger dataset from Google Places, for user-level interest analysis, as it provides a larger volume of reviews per user, essential to our high \textit{vs} low interest analysis. Originally introduced in~\cite{li_uctopic_2022,10.1145/3539618.3592036}, the dataset contains 666,324,103 geo-localized reviews of 4,963,111 venues, written by 113,643,107 users across the United States up to September 2021. Notably, 98.9\% of the data covers the period from 2015 to 2021.

\subsection{Cultural, Socioeconomic, Electoral Datasets}
\label{sec:data_socioeconomic}

To enrich the experiments, we collected diverse contextual attributes across cities in different countries. In Brazil, we used data from the 2010 Brazilian Demographic Census (IBGE)\footnote{\url{https://sidra.ibge.gov.br}}, including average monthly income, literacy rate, population size, and racial composition (White, Black, Brown, Yellow, Indigenous) for each city neighborhood. Electoral data for the 2014 Brazilian presidential election (second round) are obtained from the Regional Electoral Court (TRE)\footnote{\url{https://www.tre-pr.jus.br/eleicoes/eleicoes-anteriores/eleicoes-2014}}, following the approach of \cite{Liu2019} and \cite{Huang2023}, which uses voting discrepancies to assess political polarization. The TRE dataset provides polling place-level results across different electoral zones. Since direct neighborhood mappings are unavailable, we match voting location addresses to neighborhoods using Google Maps API, ensuring alignment with the city's limits. This method connects most polling places to neighborhoods, though some neighborhoods can lack data, mainly due to shared or misassigned polling locations, particularly in peripheral or commercial areas.

For U.S. regions, socio-demographic indicators—racial composition (White, Black, Asian, Hispanic), median household income, population size, higher education attainment, and employment rate—are obtained from the National Historical Geographic Information System (NHGIS)\cite{NHGIS}, based on the 2017–2021 American Community Survey. Political orientation is estimated using precinct-level data from the 2020 U.S. presidential election\footnote{https://github.com/TheUpshot/presidential-precinct-map-2020}, aggregated at the ZIP code and county levels. Finally, we assess cultural characteristics using Scenes Theory~\cite{culture_fingerprint}, deriving 15-dimensional scene vectors based on venue types and an alternative proxy based on venue category frequency distributions.

\subsection{Urban Areas}
\label{sec:urban_areas}

\begin{figure}[ht!]
  \centering
  \includegraphics[width=0.75\linewidth]{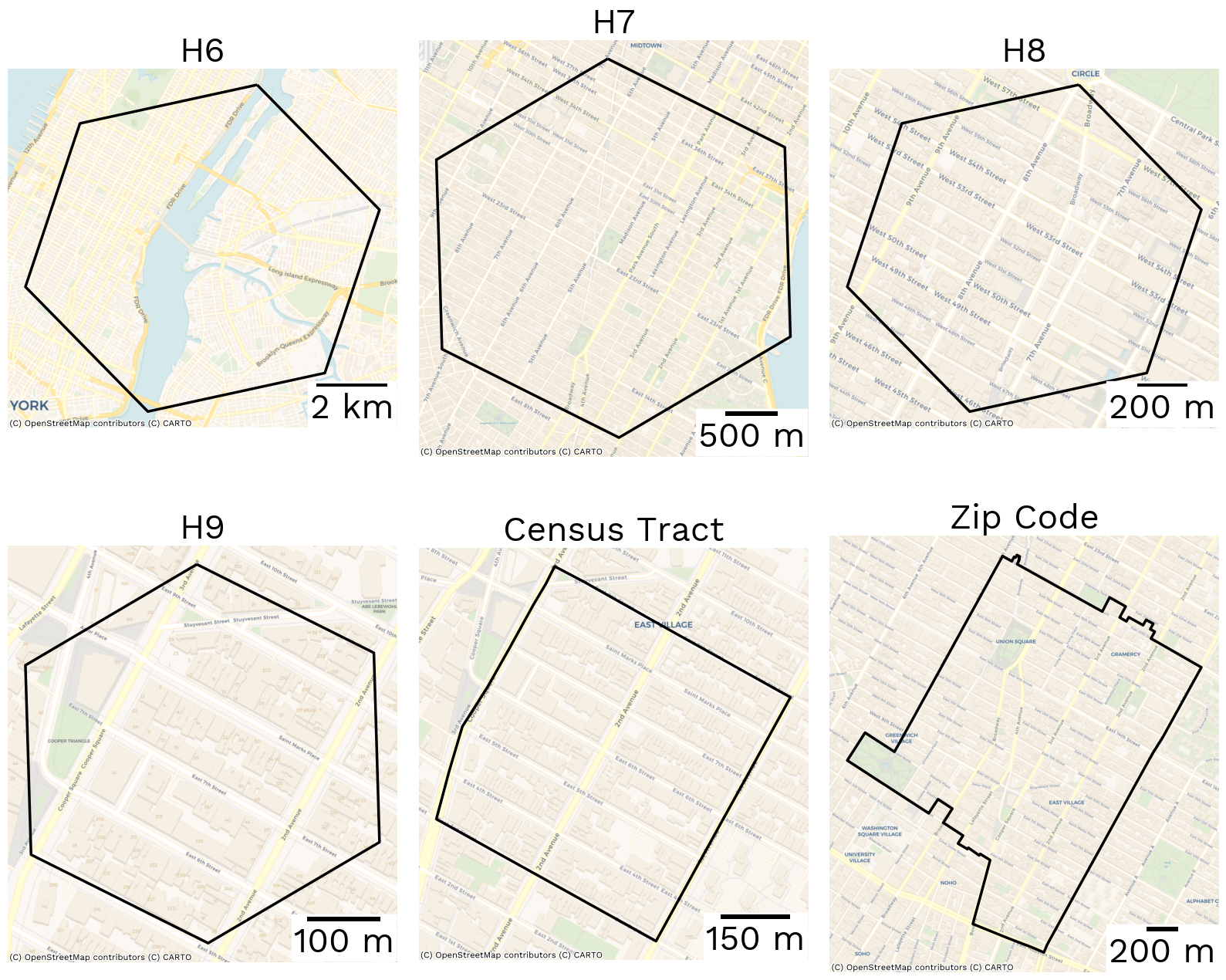}
  \caption{Different granularity levels in New York}
  \label{fig:new_york_size_comparision}
\end{figure}

For each addressed city, we use specific delimitations like census tracts and ZIP codes, neighborhoods, or boroughs.
For more coverage of different and consistent area sizes across countries, we developed h3-cities\footnote{\url{https://h3-cities.vercel.app}}, a tool that integrates OpenStreetMap with Uber’s Hexagonal Hierarchical Geospatial Indexing System to subdivide cities into hexagons of multiple resolutions. This enables consistent, multi-scalar urban analysis, as emphasized by \cite{rogov_urban_2018} and \cite{pafka_multi-scalar_2022}. We tested four different hexagonal grid resolutions ($hr$): $h6$ (average area of $36.12km^2$), $h7$ ($5.16km^2$), $h8$  ($0.74km^2$), and $h9$ ($0.11km^2$). Figure \ref{fig:new_york_size_comparision} shows a size and shape comparison of the different urban areas in New York.

\section{Methodology}

\subsection{LBSN Datasets}
In the experiments, we addressed  LBSN data from two datasets (Google Places and Foursquare) on three different areas: Curitiba, Brazil, London, UK, and several cities and counties in the USA.  Tables \ref{tab:data_description_googleplaces} and \ref{tab:data_description_foursquare} summarize the datasets from Google Places and Foursquare, respectively. Overall, the Google Places dataset offers a larger data volume in most regions, except for Curitiba, where Foursquare has more check-ins. Foursquare uses broader categories (e.g., “Food”), whereas Google Places provides more specific labels (e.g., “Italian Restaurant”). Additionally, Google Places data span a longer period, but sparse pre-2010 data have been excluded to match Foursquare’s coverage (2010–2014).

\begin{table}[ht!]
\small
\centering
\caption{Description of Google Places dataset}
\begin{tabular}{ccccc}
\bottomrule
\multirow{2}{*}{}   & {\multirow{2}{*}{Curitiba}} & {\multirow{2}{*}{London}} & \multicolumn{2}{c}{USA}                   \\ \cline{4-5} 
                    & {}                          &                         & {cities}    & counties  \\ \hline
Reviews             & {8,372}                     & {178,231}                 & {1,191,934} & 1,632,165 \\ 
Users               & {4,909}                     & {75,897}                  & {394,588}   & 486,393   \\ 
Venues              & {2,213}                     & {31,075}                  & {186,639}   & 286,075   \\ 
Categories          & {685}                       & {1,81}                    & {3,439}     & 3,793     \\ 
Period              & \multicolumn{4}{c}{from 2010 to 2014}                                                                                                             \\
\bottomrule
\end{tabular}
\label{tab:data_description_googleplaces}
\end{table}

\begin{table}[ht!]
\small
\centering
\caption{Description of Foursquare dataset}
\begin{tabular}{ccccc}
\bottomrule
\multirow{2}{*}{}   & {\multirow{2}{*}{Curitiba}} & {\multirow{2}{*}{London}} & \multicolumn{2}{c}{USA}                   \\ \cline{4-5} 
                    & {}                          &                         & {cities}    & counties  \\ \hline
Reviews             & 53,253          & 27,088  & 398,805  & 495,698  \\
Users               & 5,116           & 9,128   & 83,414   & 85,946   \\
Venues              & 8,523           & 11,104  & 122,808  & 164,655  \\
Categories          & 368             & 427     & 562      & 567      \\
Period              & 2014* & \multicolumn{3}{c}{from 2012 to 2014} \\
\bottomrule
\multicolumn{5}{l}{* from April to June}
\end{tabular}
\label{tab:data_description_foursquare}
\end{table}

\begin{figure}[h!]
\centering
    \begin{subfigure}[b]{0.5\linewidth}
      \centering
      \includegraphics[width=\linewidth]{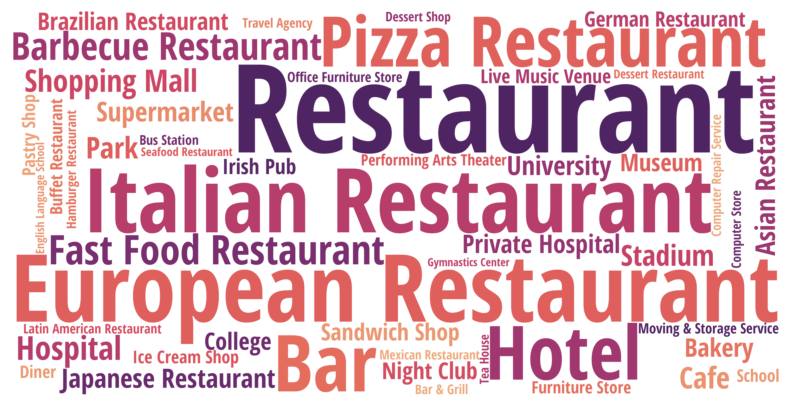}
      \caption{Google Places}
      \label{fig:google_places_wordcloud}
    \end{subfigure}%
    \begin{subfigure}[b]{0.5\linewidth}
      \centering
      \includegraphics[width=\linewidth]{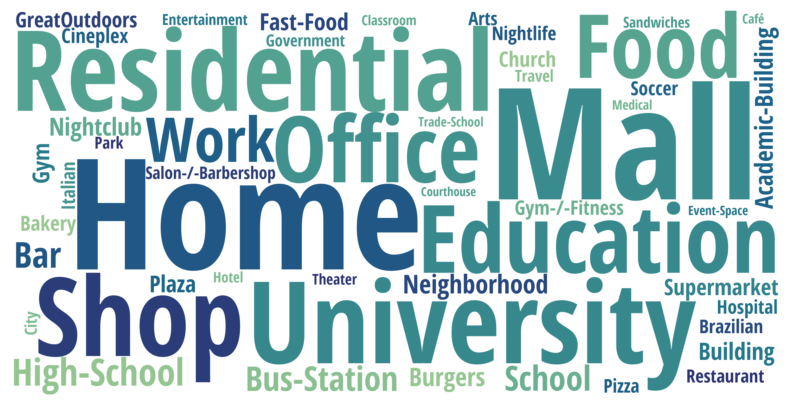}
      \caption{Foursquare}
      \label{fig:foursquare_wordcloud}
    \end{subfigure} 
    \caption{50 most popular categories in Curitiba \cite{Santos_Gubert_Delgado_Silva_2025}}
    \Description{Word Cloud of the 50 most popular categories in Curitiba}
    \label{fig:wordcloud_curitiba}
\end{figure}

A comparison of top categories in Curitiba (Figure \ref{fig:wordcloud_curitiba}) highlights key distinctions—e.g., Foursquare includes venues like homes and workplaces, which Google Places omits. 

Since our models focus on areas frequented by people, we also analyze user activity frequency and intervals between interactions. In Google Places, 25.7\% of users posted at least 2 reviews, 6.7\% posted 5, and 2.5\% posted 10 or more. Foursquare shows higher engagement: 61.3\% performed 2 check-ins, 25.3\% performed 5, and 9.6\% performed 10 or more. Regarding the time between interactions, roughly half are under 6 hours on both platforms. Longer gaps differ: one-third of Google Places users wait over a week between posts, while Foursquare users often check in every 6–24 hours (one-fourth of cases) or every 1–7 days (another fourth). In Curitiba, Foursquare users rarely go a week without checking in.
These patterns reveal that Google Places users engage less regularly, often in short bursts or after long gaps, whereas Foursquare users are more consistent. This difference and category specificity shape how each platform models urban activity.

This exploratory analysis raises a central question in our study: Can models derived from each platform yield comparable insights despite the differences? If so, may models reveal generalizable urban behavior patterns across platforms with distinct user dynamics? In the next sections, we describe how we derive the interest Networks (iNETs), the basis of our models, from LBSN data and how we explore them to provide insights into mobility patterns and user interests in complex urban environments.

\begin{figure}[h!]
  \centering
  \includegraphics[width=\linewidth]{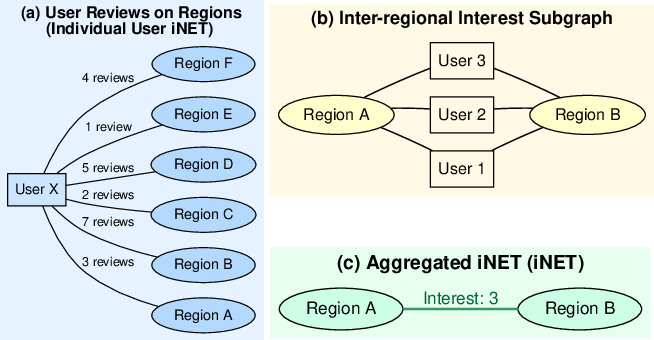}
  \caption{iNETs modelling}
  \label{fig:inets_diagram}
\end{figure}

\subsection{Formation of iNETs}
\label{sec:method_inet_definition}

iNETs model user interests in urban areas as weighted graphs. Starting from an individual user iNET (Fig.\ref{fig:inets_diagram}a), each user is linked to the regions they reviewed, weighted by the number of reviews. Regions—whether cities, counties, or neighborhoods—are of the same class. For 
every two regions we compute the inter-regional interest subgraph (Fig. \ref{fig:inets_diagram}b) as the number of users who reviewed both regions. This number becomes the edge weight of the aggregated iNET (Fig. \ref{fig:inets_diagram}c). We also define self-loops in a region as the number of users who have made at least two reviews in the same region.

\subsection{Validating iNETs across granularity levels}
\label{sec:method_validation_across_granularities}

In the second phase of our research, we assess the similarity between aggregated iNETs modeled by different LBSNs across varying granularity levels. To this end, we use the FS and GP datasets (Section~\ref{sec:LBSNs}), comparable in size and period, but with several user behavior differences as shown below, covering Curitiba (Brazil), London (UK), and the top 20 most popular U.S. cities and counties.

We apply specific metrics across multiple spatial granularities to assess the similarity between the iNETs modeled by different LBSNs. Then we analyze how area size and shape affect modeling results and their consistency. The similarity metrics used are:

\begin{itemize}
    \item \textbf{Edge weight} similarity: We calculate Pearson and Spearman correlations between corresponding edge weights in the Google Places and Foursquare networks.
    \item \textbf{Node importance} similarity: We evaluate eigenvector centrality rankings using Kendall’s Tau to measure agreement in node importance.
\end{itemize}

These metrics help assess whether different LBSNs capture similar spatial interest patterns (edges) and consistently identify key urban areas (nodes).

To address the differences observed between iNETs, we propose a new approach that defines Urban Preference Zones (UPZones) based on iNET data. By analyzing edge weights, we find that the strongest user interests typically occur between neighboring areas. This indicates that zones should be shaped by user behavior and spatial proximity rather than arbitrary boundaries.

Our approach identifies densely connected, geographically adjacent zones using the $h9$ granularity level (see an example in Figure~\ref{fig:new_york_size_comparision}) as the regions in the iNET. We detect tightly-connected communities by applying the Leiden algorithm \citep{traag_louvain_2019} ($\gamma = 1$, run until convergence). Finally, we employ TF-IDF to generate word clouds highlighting distinctive POI categories in each {\UPZones}. 
To compare urban interest zones across LBSNs, we align the Leiden-derived clusters using the Hungarian algorithm \citep{kuhn_hungarian_1955}, and evaluate their similarity through Normalized Mutual Information (NMI) for shared cluster information and Rand Index \citep{hubert_comparing_1985} for pairwise sample agreement.

\subsection{Understanding User Interests}
\label{sec:understanding_by_factors}

In the third phase, we focus on predicting if a region is highly interesting to a user based on other visited regions. Therefore, we use the GP+ dataset (see Section~\ref{sec:LBSNs}) for the larger review volume per user. We analyze the users who made reviews on at least six distinct places. We use the number of reviews to rank the places visited into low- and high-interest regions, but the ranks can blur for some users. To address the blurring of rank differences as users visit more places, particularly at the county level, the analysis focuses on a subset of users whose k-th most reviewed place has a higher review count than the (k+1)-th. For the analysis where $k=3$, this selection criterion retains 3.07M users and their 210.7M reviews, with 65,9\% of users posting all their reviews within four years and averaging 68 (median: 49) reviews each. To enhance this third phase analysis, we incorporate cultural, socioeconomic, and political contextual attributes (see Section \ref{sec:data_socioeconomic}).

Several factors are evaluated for the third research phase to understand aggregated and individual user interests. We consider geographic distance to identify mobility patterns, such as a tendency toward proximate destinations, and population size to capture preferences for residential or business-dense areas. Income and employment are used to assess economic alignment. Educational attainment, reflected in the proportion of residents holding bachelor’s degrees or literacy (in Curitiba), proxies for intellectual engagement. Racial composition reveals demographic affinity. Political orientation provides insight into the role of ideological alignment. Venue categories evaluate the similarity of the experiences that the places offer. Scenes help us understand the importance of cultural aspects in urban areas. 

We used several distance and similarity metrics to better assess how regions resemble each other based on each characteristic.

On the aggregated level of iNETs, we correlate the edge weights with the distance and similarity metrics for each feature analyzed.
For the individual user level of iNETs, we propose a method for personalized user interest modeling and recommendation. We first differentiate a user's visited regions (e.g., cities, neighborhoods) into high-interest (top) and low-interest (bottom) sets based on review counts. Specifically, we rank regions by review frequency and select the top k as the high-interest set (e.g., for six visited cities and $k=3$, the top 3 by reviews are high-interest, the rest are low-interest). This can be further applied at multiple scales. For instance, using $m=2$, we could identify the top 2 most-reviewed ZIP  codes within each top k city to form a finer-grained high-interest set. 

We then train a LightGBM classifier to predict whether a region belongs to the user's high-interest set, based on the mean distances/similarities to the other top and bottom regions. To understand the factors driving high interest, we analyze feature importance using permutation importance and SHAP values \cite{NIPS2017_7062} on the test set (derived from an 80/20 train-test split).

The model assigns greater weight to high-interest regions, based on the ratio of low- to high-interest samples in the training data. Hyperparameters are tuned using a randomized grid search.

Finally, we show the power of this modeling by analyzing the 'returners' vs. 'explorers' dichotomy \cite{Pappalardo2015}. Returners are identified as users where the radius of gyration of their top k locations exceeds half the radius of gyration of all their visited locations, demonstrating the method's ability to capture distinct behavioral patterns.

\section{Results}
\label{secResults}

\subsection{The impact of different Granularity Levels and Platforms}

\begin{figure}[h!]
  \centering
  \includegraphics[width=\linewidth]{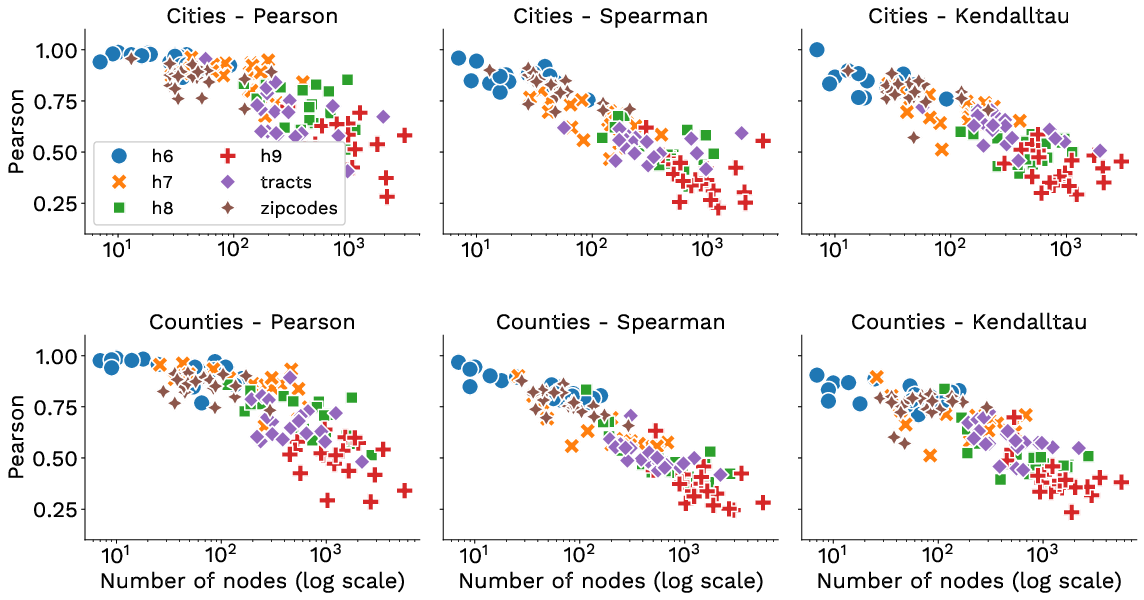}
  \caption{Results of correlation between both iNETs in USA}
  \label{fig:correlations_top_20}
\end{figure}

In the U.S., we modeled 240 iNETs per LBSN (GP or FS), i.e., 20 cities (or counties), each evaluated using six granularities. The smallest is New York County (GP, h6): 7 nodes, 28 edges. The largest is Los Angeles County (GP, h9): 10,363 nodes, 455,976 edges. On all these iNETs, we perform similarity analyses between LBSNs (see Section \ref{sec:method_validation_across_granularities}). 
The results are presented in Figure~\ref{fig:correlations_top_20}. 

In this figure, the y-axis indicates the correlation value analyzed for each granularity level, while the x-axis presents, on a logarithmic scale, the number of nodes in both iNETs (GP and FS). Observing Figure~\ref{fig:correlations_top_20}, there is an evident pattern that the more nodes an iNET has, the greater the differences between the iNETs modeled by the studied LBSNs. It is also important to highlight that counties vary in area from  $87$ km$^2$ to $23,895$ km$^2$ while cities vary from $177$ km$^2$ to $1,740$ km$^2$. As a result, some regions have more nodes at the  $h7$ granularity level than other regions at the $h8$ level, even if the $h8$ nodes are smaller than the $h7$ nodes. 

\begin{figure}[h!]
  \centering
  \includegraphics[width=\linewidth]{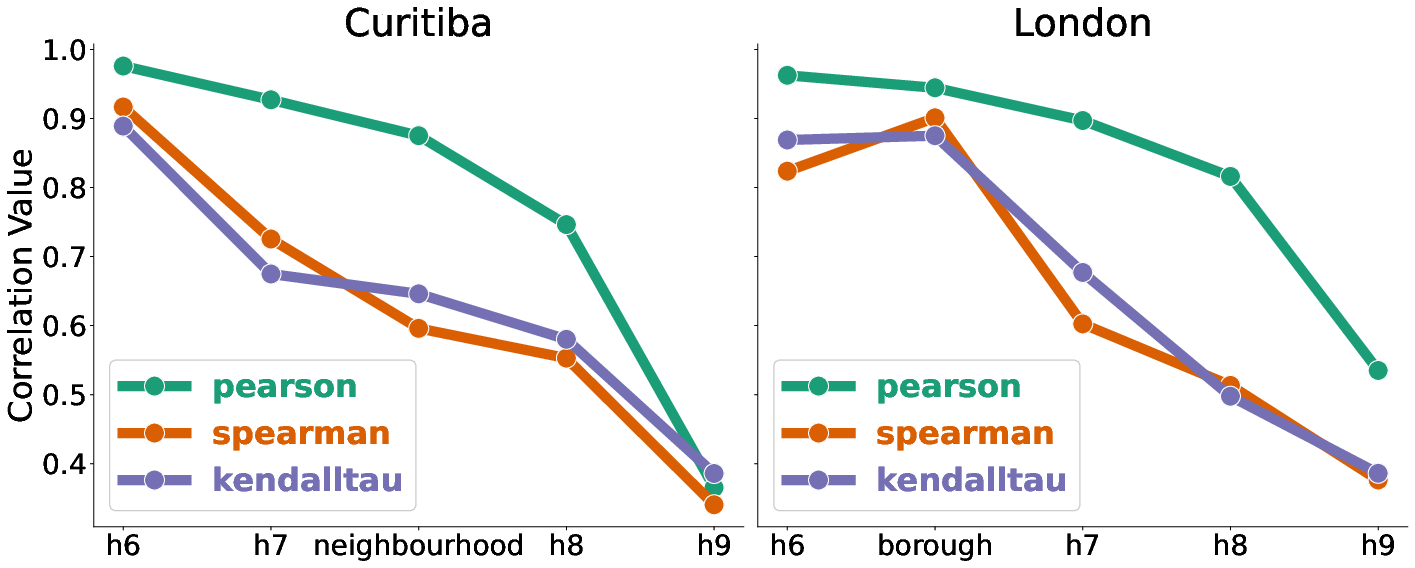}
c  \caption{Results of correlation between both iNETs}
  \label{fig:curitiba_and_london_inets_correlations}
\end{figure}

Figure~\ref{fig:curitiba_and_london_inets_correlations} presents similarity correlations between iNETs from Google Places and Foursquare in Curitiba, segmented by neighborhoods, London segmented by boroughs, and both using $h6$ to $h9$ granularities. Notably, neighborhood sizes in Curitiba typically fall between $h7$ and $h8$, and London boroughs fall between $h6$ and $h7$. The findings suggest that finer spatial granularity reveals greater differences between iNETs.

The granularity level h6 ($36.12$ km$^2$) provides consistent insights across LBSNs when examining central urban areas and their broader relationships. Finer granularities (h7, h8) yield more detailed analyses but may introduce diverse iNETs, capturing LBSN-specific behaviors rather than general trends. The h9 level, while enabling sharper differentiation between LBSNs, suffers from increased noise due to sparse data and higher node counts.

Government-defined divisions (e.g., census tracts, ZIP codes, neighborhoods) align with certain granularities: census tracts and neighborhoods resemble h7/h8, while ZIP codes and boroughs parallel h6 in homogeneity. Socioeconomic data linked to administrative divisions aid in investigating interest factors, but hierarchical granularities offer flexibility, from broad patterns (h6) to micro-scale interactions (h9). These findings align with \cite{wu_multi-scale_2020}, confirming that granularity choice shapes complementary urban insights.

\subsection{{\UPZonesTitle} ({\UPZonesShort}s)}

This section uses London as a case study for the \UPZones, although similar results are found in Curitiba. In London (using h9 cells), GP iNET includes 123,292 reviews from 67,276 users, forming 396,876 edges (1,533 self-loops) across 5,654 of 6,600~cells. FS iNET has 22,738 check-ins from 8,178 users, yielding 34,075 edges (1,209 self-loops) across 2,899 cells.

Using the method from Section~\ref{sec:method_validation_across_granularities}, we identify 1,760 {\UPZonesShort}s from Google Places and 1,023 from Foursquare. Comparing these zones to the city's boroughs \cite{Santos_Gubert_Delgado_Silva_2025} shows that {\UPZones} offer finer-grained insights into user interests than traditional administrative divisions.

\begin{figure}[h!]
      \centering
      \includegraphics[width=\linewidth]{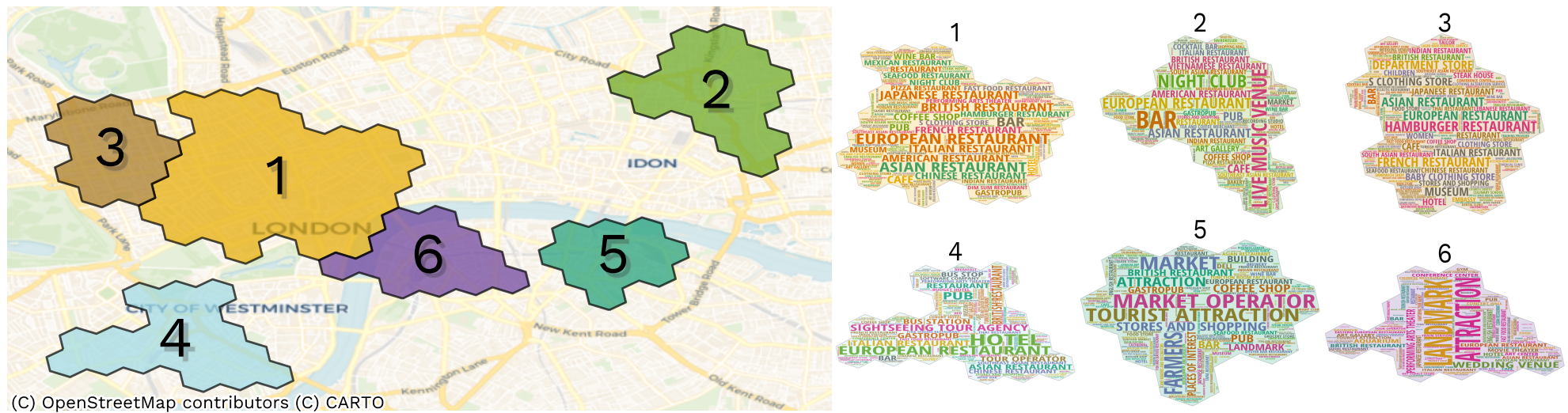}
      \caption{Characterization of the main {\UPZonesShort}s found}
      \label{fig:london_areas_description}
\end{figure}

Figure~\ref{fig:london_areas_description} shows how {\UPZonesShort}s constructed from the iNET at $h9$ granularity capture semantically meaningful areas in London. Each colored and numbered region represents a {\UPZonesShort} derived from Google Places, with numbers indicating the top 6 zones by activity. Word clouds illustrate the most prominent categories in each zone (see Section~\ref{sec:method_validation_across_granularities}), emphasizing the dominant place types in these areas.

The figure shows that {\UPZonesShort}~1 includes Soho, a vibrant area known for nightlife and dining; {\UPZonesShort}~2 captures an artistic hub hosting performances and creative activity; and {\UPZonesShort}~3 corresponds to Mayfair, noted for high-end retail. {\UPZonesShort}~6 covers the South Bank, home to cinemas, galleries, and landmarks like the London Eye. Other {\UPZonesShort}s similarly align with popular areas frequented by users with shared demographic interests. While Foursquare data also identifies central zones, its smaller dataset may limit comparability. Overall, the proposed method effectively detects {\UPZones}, revealing areas that transcend administrative boundaries and offer deeper insight into urban user behavior.

Using the metrics from Section \ref{sec:method_validation_across_granularities}, we compare {\UPZonesShort}s derived from Google Places and Foursquare iNETs, obtaining an NMI of 0.6364 and a Rand Index of 0.7378. These results suggest that, despite differences in the underlying iNETs, both LBSNs yield moderately similar {\UPZonesShort}s, supporting the robustness of our modeling approach. 

\subsection{Influence of Cultural, Socioeconomic, Political, and Geographic Factors}

For the aggregate level iNET for Google Places users in Curitiba, we analyze 1,127 users (with reviews $\geq 2$) and 4,590 reviews, while Foursquare's iNET encompasses 3,933 users and 52,033 check-ins. The resulting Google Places graph covers 69 of 75 neighborhoods, with 1,287 edges (53 self-loops), while Foursquare's has 75 nodes and 1,618 edges (58 self-loops). Both networks show concentrated activity in central commercial and leisure districts, with Google Places extending further into peripheral areas. We use the neighborhoods due to their association with socioeconomic data.

As per Section \ref{sec:method_validation_across_granularities}, we correlate the edge weights of the modeled networks with the pairwise node distances across key characteristics for filtered iNETs. Using Spearman correlation, the results for Google Places and Foursquare networks are shown in Table \ref{tab:spearman_correlations_curitiba_inets}. We
kept only the top 75\% of edges by weight,
emphasizing stronger interest connections.

\begin{table}[ht]
\centering
\caption{Spearman Correlations with the Analyzed Factors in Curitiba}
\begin{tabular}{lcc}
\toprule
 & Google Places & Foursquare \\
\midrule
Population & -0.14 & -0.03 \\
Income Differences & -0.17 & -0.06 \\
Cultural Affinity & -0.28 & 0.00 \\
Political Polarization & -0.26 & -0.14 \\
Venues Similarity & 0.33 & 0.12 \\
Racial Composition & -0.26 & -0.14 \\
Education Levels & -0.33 & -0.23 \\
Geographic Distance & -0.4 & -0.42 \\
\bottomrule
\end{tabular}
\label{tab:spearman_correlations_curitiba_inets}
\end{table}

The analysis shows that geographic distance has the most strongly negative correlation (Table \ref{tab:spearman_correlations_curitiba_inets}), reflecting known mobility constraints where long-distance visits are less common \citep{Cheng2021, Gonzlez2008, Rhee2008, Brockmann2006}. Then, education levels and racial composition showed a lower but interesting correlation in Curitiba. After that, venue similarity emerges as the second strongest correlated in Google Places, highlighting the relevance of shared venue experience traits in driving urban interest, but lower in Foursquare, likely due to having categories like the user's own home. Finally, political polarization, cultural affinity (Except by GP, where a more venue-centered data better fits with Scenes Theory), income differences, and population size exhibited weak correlations with people’s interest patterns, suggesting users do not strongly prefer areas with similar or dissimilar aspects based on these traits.

\begin{figure}[h!]
      \centering
      \includegraphics[width=\linewidth]{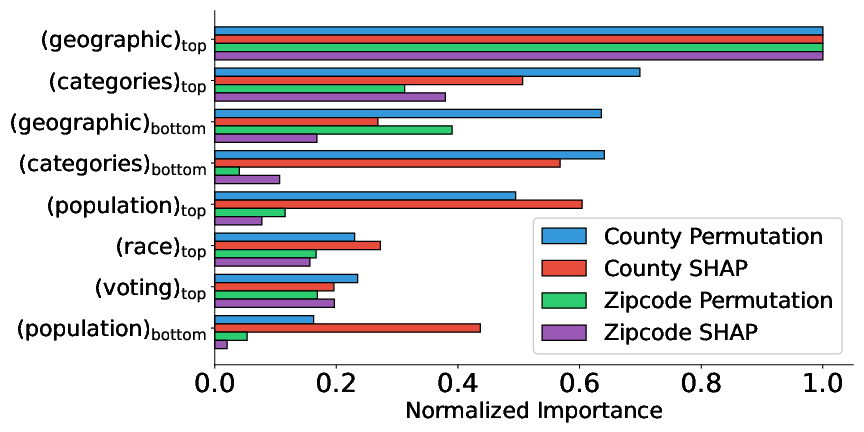}
      \caption{Most influential features for high-interest areas}
      \label{fig:model_feature_importance_user_level}
\end{figure}

For the user-level analysis on U.S. counties (see Section \ref{sec:understanding_by_factors}), we perform a preliminary test using $k=3$, which yields a recall of 0.68 and an F1-score of 0.55 on the test set. When applied to the corresponding ZIP codes using $m=2$, the model achieves a recall and F1-score of 0.71. To identify the most influential features for distinguishing high-interest areas, we use permutation feature importance and mean absolute SHAP values. After applying min-max normalization, the top 10 features—ranked by their mean normalized importance across scenarios—are presented in Figure \ref{fig:model_feature_importance_user_level}.

Geographic distance to top regions is the strongest predictor, indicating that users favor venues near those they’ve previously visited more. Similarity in venue categories to top regions helps identify high-interest areas, while dissimilarity suggests low-interest zones. Population size primarily influences preferences at the county level, reflecting urban vs. rural appeal. Lastly, race and political alignment show moderate importance at both the county and ZIP code levels.

Analyzing the dichotomy between the returners and explorers reveals a notable pattern. When predicting regions of interest separated for these classes of users, the model considering explorers outperforms the one focusing on returners, achieving a recall of 0.80 and an F1-score of 0.63, compared to returners’ 0.71 recall and 0.54 F1-score. For $k=3$, 47\% of users are classified as explorers. The key difference between groups lies in feature importance. For explorers, proximity to other high-interest cities is the most influential factor—nearly twice as important as the second-ranked feature, venue similarity. In contrast, returners are more influenced by venue category similarity, with geographic distance playing a minor role. This distinction appears only at the city level; at the neighborhood level, both groups show similar behavior in selecting high-interest areas.

To better demonstrate the strength of our user-level modeling, we propose the CityHood demo\footnote{\url{https://cityhood.vercel.app/}}, which allows users to explore personalized city and neighborhood recommendations based on their past preferences, enhanced with AI-powered explainability.

\section{Conclusion}
\label{secConclusion}



This work introduces \textit{Interest Networks (iNETs)} to model urban preferences using Location-Based Social Network (LBSN) data from Google Places and Foursquare. Our analysis reveals that coarser spatial granularities (e.g., city or borough levels) yield consistent cross-platform insights, while finer resolutions (e.g., neighborhoods or hexagons) expose platform-specific behaviors. Key findings highlight that, in general, geographic proximity and venue similarity dominate user interest patterns, while socioeconomic and political factors have a comparatively smaller influence.

To bridge Google Places and Foursquare platform discrepancies, we propose \textit{Urban Preference Zones (UPZones)}---behaviorally coherent clusters that transcend administrative boundaries. 
We further develop a scalable, explainable recommendation system that adapts to user profiles (e.g., \textit{explorers} vs. \textit{returners}) and provides interpretable justifications via XAI. Although evaluating such a system is challenging—e.g., varying $k$ and $m$ or testing its predictions on various subsets and permutations of user history—we plan to explore this further in future work. Our public demo demonstrates practical applications for urban planning and mobility research.

While the study uses partially dated datasets, the methodology remains applicable to modern platforms and contexts. In future work, we aim to better investigate how the sampling biases in the LBSNs studied impact the results, integrate textual reviews through natural language processing, incorporate time-aware mobility patterns, and validate our findings through local expert feedback. Our tools, datasets, and public demo provide a foundation for further research and applications in urban computing, mobility modeling, and recommender systems.

\begin{acks}
This research is partially supported by the SocialNet project (grant 2023/00148-0 of FAPESP), by CNPq (grants 314603/2023-9, 441444/2023-7, 409669/2024-5, and 444724/2024-9) and the INCTs ICoNIoT (CNPq grant 405940/2022-0 and CAPES FinanceCode 88887.954253/2024-00) and TILD-IAR (CNPq grant 408490/2024-1).
\end{acks}

\bibliographystyle{ACM-Reference-Format}
\bibliography{bibliography}

\newpage

\appendix

\section{Additional Reproducibility Details}

Besides the details in the methodology section, we present some information on the implementation to aid in reproducibility tasks.

Census tract and ZIP code delimitations have been obtained from the U.S. Census Bureau's Tiger geographic database\footnote{\url{https://www.census.gov/cgi-bin/geo/shapefiles/index.php}}; and the neighborhood polygons from the Curitiba city hall website\footnote{\url{https://ippuc.org.br/geodownloads/geo.htm}}. London's boroughs were collected from the London Datastore\footnote{\url{https://data.london.gov.uk/dataset/statistical-gis-boundary-files-london}}. 

The h3-cities\footnote{\url{https://h3-cities.vercel.app}} tool makes it easy to retrieve H3 cells. Enter the name of your desired city or region, select the resolution, and the download option will appear. You'll also see a visual preview of the area and its corresponding hexagons.

We use several distance and similarity metrics between regions, where each region belongs to the same granularity class. The metrics are as follows:

Geographic Proximity is computed as the Euclidean distance between the centroids of two regions in the ESRI:102005 (USA) and UTM22S (Curitiba) projections.

Population, income, education (bachelor or literacy rates), employment, and voting similarities were measured as the absolute difference between values in each region.

Race differences were computed as:
$$\frac{1}{2}\sum_{i=1}^n \left| \frac{P_i(r)}{P(r)} - \frac{P_i(r')}{P(r')} \right|$$ where $P(r)$ represents the population size of region $r$, and $P_i(r)$ is the population size of the i-th racial category in region $r$. This work uses the racial categories: White, Black, Asian, and Hispanic for the USA regions and White, Black, Brown, Yellow, and Indigenous for Brazil.

Cultural Similarity was calculated using the Euclidean Distance in the 15 scenes dimensions studied.

Venue similarity was computed using cosine similarity between each region's frequency vectors of venue categories.

On the training of the LightGBM model, the following ranges of hyperparameter are tested on the random grid search: The hyperparameter were tested over the following ranges: number of trees (100–300), learning rate (0.01–0.1), maximum tree depth (3–10), number of leaves per tree (20–50), minimum child samples (10–30), and L1/L2 regularization strengths (0–0.5).

To ensure consistency in filtering the top 75\% of edges by weight, we first sorted all edges in descending order of weight. For edges with weights equal to the cutoff threshold, we broke ties deterministically by sorting lexicographically using node IDs (neighborhood names), with smaller nodes first. This guarantees that the same subset of edges is selected across runs.

All analyses were run on a server with an Intel i7-10700 CPU and 128 GB RAM. After obtaining the geo-localized reviews, iNETs were built using DuckDB, which enabled efficient computation of region-to-region interest graphs through SQL-based aggregation over user co-visitation data. By joining check-in or review records on user IDs and grouping by region pairs, we calculated the number of users who interacted with both regions, forming the weighted edges of the iNETs. This method was applied consistently across all spatial granularities ($h6$–$h9$, census tracts, ZIP codes, neighborhoods, and boroughs) and for both LBSNs—Google Places (GP) and Foursquare (FS). Thanks to DuckDB’s optimized performance for analytical queries, the complete construction of all iNETs across all cities and granularities, for both GP and FS datasets, as well as the correlation analyses, was completed in under 5 minutes total. The construction of Urban Preference Zones (UPZones) in London using the Leiden algorithm on $h9$ granularity took about 1s, with parameters $\gamma = 1$, $\text{n\_iterations} =-1$ (running until convergence), and using the iNET edge weights. LightGBM training, including SHAP and permutation importance analysis, required approximately 2 hours each for both city-level and ZIP code–level models.


\end{document}